\documentstyle[a4,11pt]{article}
\begin{document}

\begin{titlepage}
\vskip 2cm
\begin{flushright}
Preprint CNLP-1995-07
\end{flushright}
\vskip 2cm
\begin{center}
{\large {\bf SURFACE, CURVES AND THE LAKSHMANAN EQUIVALENT COUNTERPARTS
OF THE SOME MYRZAKULOV EQUATIONS}}\footnote{Preprint
CNLP-1995-07.Alma-Ata.1995 \\
cnlpmyra@satsun.sci.kz}
\vskip 2cm

{\bf G.N.Nugmanova  }

\end{center}
\vskip 1cm

$^{a}$ Centre for Nonlinear Problems, PO Box 30, 480035, Almaty-35, Kazakstan

\begin{abstract}
The Lakshmanan equivalent counterparts of the some Myrzakulov equations
are found.
\end{abstract}


\end{titlepage}

\setcounter{page}{1}
\newpage

{\bf I.INTRODUCTION}

In a series of papers [1] were presented the new class integrable
and nonintegrale spin equations
and the unit spin description of soliton equations.  In this paper using
the geometrical method we will find the Lakshmanan equivalent[2] (according
to the terminology of ref.[1])  counterparts of the some Myrzakulov equations.

Consider a n-dimensional space with the basic unit vectors: $\vec e_{1}=
\vec S, \vec e_{2}, ... ,\vec e_{n}$. Then the M-0 equation has the form[1]
$$
\vec S_{t} = \sum^{n}_{i=2} a_{i} \vec e_{i}          \eqno(1)
$$
where $a_{i}$ are real functions, $\vec S = (S_{1}, S_{2}, ... , S_{n}),
\vec S^{2} = E = \pm 1$. This equation admit the many interesting class
integrable and nonintegrable reductions[1]. Below we will find the
L-equivalent counterparts of the some integrable reductions and
only for the cases $n = 2, 3$. To this end, starting from the
results of the ref.[8] we consider the motion of surface in
the 3-dimensional space which generated by a position vector
$\vec r(x,y,t) = r(x^{1}, x^{2}, t)$. According to the C-approach from
the ref.[1], the main elements of which we present in this section,
 let x and y are local coordinates on the surface. The
first and second fundamental forms in the usual notation are given by
$$ I=d\vec r d\vec r=Edx^2+2Fdxdy+Gdy^2 \eqno (2) $$
$$ II=-d\vec r d\vec n=Ldx^2+2Mdxdy+Ndy^2 \eqno (3) $$
where $$ E=\vec r_x\vec r_x=g_{11}, F=\vec r_x\vec r_y=g_{12},
G=\vec r_{y^2}=g_{22},   $$
$$ L=\vec n\vec r_{xx}=b_{11}, M=\vec n\vec r_{xy}=b_{12},
N=\vec n\vec r_{yy}=b_{22}, \vec n=\frac{(\vec r_x \wedge \vec r_y)}
{|\vec r_x \wedge \vec r_y|}  $$

In the C-approach[1], the starting set of equations reads as
$$ \vec r_{t} = a \vec r_x + b \vec r_y + c \vec n \eqno (4a) $$
$$ \vec r_{xx}=\Gamma^1_{11} \vec r_x + \Gamma^2_{11} \vec r_y +L \vec n \eqno (4b) $$
$$ \vec r_{xy}=\Gamma^1_{12} \vec r_x + \Gamma^2_{12} \vec r_y +M \vec n \eqno (4c)$$
$$ \vec r_{yy}=\Gamma^1_{22} \vec r_x + \Gamma^2_{22} \vec r_y + N \vec n \eqno (4d)$$
$$ \vec n_x=p_1 \vec r_x+p_2 \vec r_y \eqno (4e) $$
$$ \vec n_y=q_1 \vec r_x+q_2 \vec r_y \eqno (4f) $$
where $ \Gamma^k_{ij} $ are the Christoffel symbols of the second kind defined by
the metric $ g_{ij} $ and $ g^{ij}=(g_{ij})^{-1} $ as
$$ \Gamma^k_{ij}=\frac{1}{2} g^{kl}(\frac{\partial g_{lj}} {\partial x^i}+
   \frac {\partial g_{il}}{ \partial x^j}-\frac{\partial g_{ij}}
   {\partial x^l}) \eqno (5) $$
The coefficients $ p_i, q_i $ are given by
$$ p_i=-b_{1j}g^{ji}, \,\,\, q_i=-b_{2j}g^{ji} \eqno (6) $$
The compatibility conditions $ \vec r_{xxy}=\vec r_{xyx} $ and
$ \vec r_{yyx}=\vec r_{xyy} $ yield the following Mainardi-Peterson-Codazzi
equations (MPCE)
$$ R^l_{ijk} = b_{ij}b^l_{k}-b_{ik}b^l_{j} \eqno (7a) $$
$$ \frac{\partial b_{ij}}{\partial x^k}-\frac{\partial b_{ik}}{\partial
    x^j}=\Gamma^s_{ik}b_{is}-\Gamma^s_{ij}b_{ks}
\eqno (7b) $$
where $ b^j_i=g^{jl}b_{il} $ and the curvature tenzor has the form
$$ R^l_{ijk} = \frac{\partial \Gamma^l_{ij}}{\partial x^k}-\frac{\partial \Gamma^l_{ik}}
{\partial x^j}+\Gamma^s_{ij} \Gamma^l_{ks}-\Gamma^s_{ik} \Gamma^l_{js}
\eqno (8) $$

Let
$\vec Z = (\vec r_{x}, \vec r_{y}, \vec n)^{t}$ . Then
$$ \vec Z_{x} = A \vec Z \eqno (9) $$
$$ \vec Z_{y} = B \vec Z \eqno (10) $$
where
\begin{displaymath}
A =
\left ( \begin{array}{ccc}
\Gamma^{1}_{11} & \Gamma^{2}_{11} & L \\
\Gamma^{1}_{12} & \Gamma^{2}_{12} & M \\
p_{1}           & p_{2}           & 0
\end{array} \right) ,
\end{displaymath}
\begin{displaymath}
B =
\left ( \begin{array}{ccc}
\Gamma^{1}_{12} & \Gamma^{2}_{12} & M \\
\Gamma^{1}_{22} & \Gamma^{2}_{22} & N \\
q_{1}           & q_{2}           & 0
\end{array} \right)
\end{displaymath}
Hence we get the new form of the MPCE(7)
$$ A_y - B_x + [A, B] = 0 \eqno (11) $$

Let us introduce the orthogonal trihedral[1]
$$ \vec e_{1} = \frac{\vec r_x}{\surd E}, \,\,\,
\vec e_{2} = \frac{\vec r_y}{\surd F}, \,\,\, \vec e_{3} = \vec e_{1} \wedge
\vec e_{2} = \vec n    \eqno(12a) $$
or
$$ \vec e_{1} = \frac{\vec r_x}{\surd E}, \,\,\,
\vec e_{2} = \vec n, \,\,\, \vec e_{3} = \vec e_{1} \wedge
\vec e_{2}    \eqno(12b) $$

Let $ \vec r_x^2 = E = \pm 1,  \,\,F = \vec r_{x}
\vec r_{y} = 0$.  Then from the previouse equations after some algebra
we get[1]
\begin{displaymath}
\left ( \begin{array}{ccc}
\vec e_{1} \\
\vec e_{2} \\
\vec e_{3}
\end{array} \right)_{x}
\end{displaymath}

$$
= C
\left ( \begin{array}{ccc}
\vec e_{1} \\
\vec e_{2} \\
\vec e_{3}
\end{array} \right)  \eqno(13a)
$$

$$
\left ( \begin{array}{ccc}
\vec e_{1} \\
\vec e_{2} \\
\vec e_{3}
\end{array} \right)_{y} = D
\left ( \begin{array}{ccc}
\vec e_{1} \\
\vec e_{2} \\
\vec e_{3}
\end{array} \right) \eqno(13b)
$$

Here
\begin{displaymath}
C =
\left ( \begin{array}{ccc}
0   & k     & 0 \\
-Ek & 0     & \tau  \\
0   & -\tau & 0
\end{array} \right) ,
\end{displaymath}
\begin{displaymath}
D =
\left ( \begin{array}{ccc}
0       & m_{3}  & -m_{2} \\
-Em_{3} & 0      & m_{1} \\
Em_{2}  & -m_{1} & 0
\end{array} \right)
\end{displaymath}
Now the MPCE (7) and/or (11) becomes
$$C_y - D_x + [C, D] = 0 \eqno (14) $$
Hence we obtain
$$ (m_{1}, m_{2}, m_{3}) =
(\partial ^{-1}_{x}(\tau_{y} + k m_{2}), m_{2},
\partial ^{-1}_{x}(k_{y} - \tau m_{2}) ) \eqno(15a) $$
$$ m_{2} = \partial ^{-1}_{x}(\tau m_{3} - k m_{1}) \eqno(15b) $$

The time evolution of $\vec e_{i}$ we can write in the form[1]
$$
\left ( \begin{array}{ccc}
\vec e_{1} \\
\vec e_{2} \\
\vec e_{3}
\end{array} \right)_{t} = G
\left ( \begin{array}{ccc}
\vec e_{1} \\
\vec e_{2} \\
\vec e_{3}
\end{array} \right) \eqno(16)
$$
with
\begin{displaymath}
G =
\left ( \begin{array}{ccc}
0       & \omega_{3}  & -\omega_{2} \\
-E\omega_{3} & 0      & \omega_{1} \\
E\omega_{2}  & -\omega_{1} & 0
\end{array} \right) ,
\end{displaymath}
So, we have
$$C_t - G_x + [C, G] = 0 \eqno (17a) $$
$$D_t - G_y + [D, G] = 0 \eqno (17b) $$
\\
\\
\\
\begin{center}
{\bf II. THE L - EQUIVALENT COUNTERPARTS OF THE
SOME MYRZAKULOV EQUATIONS:
the 2-dimensional case}
\end{center}

In this case the M-0 equation(1) becomes
$$
\vec S_{t} = a_{2} \vec e_{2}          \eqno(18)
$$
and  $\vec S = (S_{1}, S_{2}),
\vec S^{2} = E = \pm 1, \tau = c =0$.
So, we have[1]
$$
\left ( \begin{array}{ccc}
\vec e_{1} \\
\vec e_{2}
\end{array} \right)_{x} = C
\left ( \begin{array}{ccc}
\vec e_{1} \\
\vec e_{2}
\end{array} \right)  \eqno(19a)
$$

$$
\left ( \begin{array}{ccc}
\vec e_{1} \\
\vec e_{2}
\end{array} \right)_{y} = D
\left ( \begin{array}{ccc}
\vec e_{1} \\
\vec e_{2}
\end{array} \right) \eqno(19b)
$$
Here
\begin{displaymath}
C =
\left ( \begin{array}{ccc}
0   & k     \\
-Ek & 0
\end{array} \right) ,
\end{displaymath}
\begin{displaymath}
D =
\left ( \begin{array}{ccc}
0       & m_{3} \\
-Em_{3} & 0
\end{array} \right)
\end{displaymath}
Now from the MPCE (14)  we obtain
$$  m_{3} =
\partial ^{-1}_{x}k_{y} \eqno(20) $$
For the time evolution we get[1]
$$
\left ( \begin{array}{ccc}
\vec e_{1} \\
\vec e_{2}
\end{array} \right)_{t} = G
\left ( \begin{array}{ccc}
\vec e_{1} \\
\vec e_{2}
\end{array} \right) \eqno(21)
$$
where
$$
G =
\left ( \begin{array}{ccc}
0       & \omega_{3}  \\
-E\omega_{3} & 0
\end{array} \right).
$$

Now already we ready using this formalism  to construct the L-equivalent
counterparts of the some Myrzakulov equations.
\\
\\
{\bf Examples}
\\

1) The M-IV equation has the following L-equivalent counterpart
$$ \phi_t+\phi_{xxy}-3E(\phi\partial^{-1}_x(\phi^{2})_y)_x=0 \eqno (22a) $$
which is the 2+1 dimensional mKdV[9].

2) The M-XXI equation has the following L-equivalent counterpart
$$ \phi_t+\phi_{xxy}-3E(\phi\partial^{-1}_x\phi_y)_x=0 \eqno (22b) $$
which is the 2+1 dimensional KdV[9].

\begin{center}
{\bf III. THE L - EQUIVALENT COUNTERPARTS OF THE
SOME MYRZAKULOV EQUATIONS: the 3-dimensional case}
\end{center}

In this case work equations(12)-(16) and the M-0 equation becomes
$$
\vec S_{t} = a_{2} \vec e_{2} +a_{3}\vec e_{3}
$$
and  $\vec S = (S_{1}, S_{2}, S_{3}),
\vec S^{2} = E = \pm 1$. Using these equations we construct
the L-equivalent counterparts of the some Myrzakulov and Ishimori
equations[1].  Below we present  only the final results.
\\
{\bf Examples}
\\

1) The Myrzakulov-III equation
$$ \vec S_{t}=(\vec S\wedge \vec S_{y}+u\vec S)_x+2b(cb+d)\vec S_{y}
     -4cv\vec S_{x} \eqno (23a) $$
$$ u_x=-\vec S_(\vec S_{x}\wedge \vec S_{y}), v_x=\frac{1}{4(2bc+d)^2}
   (\vec S^2_{1x})_y \eqno (23b) $$
in this case
$$ (m_1, m_2, m_3)=(\partial^{-1}_x(\tau_y+km_2), -\frac{-u_x}{k},
    \partial^{-1}_x(k_y-\tau m_2))  \eqno (24) $$
and the L-equivalent is the following set of equations [1]
$$ i\phi_t=\phi_{xy}-4ic(V\phi)_x+2d^2V\phi, V_x=2E(|\phi|^2)_y. \eqno (25) $$

Note that equations (23) and (25) admit the following integrable
reductions: a) the M-I[1] and the
Zakharov[4] equations, as c=0; b) the M-II[1] and Strachan[9] equations, as d=0,
respectively [1].

2) The Myrzakulov-VIII equation looks like[1]
$$ iS_t=\frac{1}{2}[S_{xx},S]+iu_xS_x   \eqno (26a) $$
$$ u_{xy}=\frac{1}{4i}tr(S[S_y,S_x])  \eqno (26b) $$
where the subscripts denote partial derivatives and S denotes the spin
matrix $ (r^2=\pm1)$

$$S= \pmatrix{
S_3 & rS^- \cr
rS^+ & -S_3
}, \eqno (27)$$
$$ S^2=I    $$

Equations (26) are integrable, i.e. admits Lax representation
and different type soliton solutions [1]. The Lakshmanan equivalent
counterpart of the M-VIII equation (26) has the form[1]
$$ iq_t + q_{xx} + v q = 0,  \eqno(28a) $$
$$ ip_t - q_{xx} - v q = 0,\eqno(28b) $$
$$ v_y=2(pq)_x \eqno(28c) $$
where $p = E\bar q$. On the other hand, in [5] was shown that eqs.(26)
and (28) are gauge equivalent each other.

3) The Ishimori equation
$$ iS_t+\frac{1}{2}[S,M_{10}S]+A_{20}S_x+A_{10}S_y = 0 \eqno(29a)$$
$$ M_{20}u=\frac{\alpha}{4i}tr(S[S_y,S_x]) \eqno(29b)$$
where $ \alpha,b,a  $= consts and
$$ M_{j0} = M_{j},\,\,\, A_{j0}=A_{j}\,\,\,\, as \,\,\,\,a = b = -\frac{1}{2}. $$

The L-equivalent counterpart has the form [1]
$$ iq_t+M_{10}q+vq=0 \eqno(30a)$$
$$ ip_t-M_{10}p-vp=0 \eqno(30b)$$
$$ M_{20}v=M_{10}(pq) \eqno(30c)$$
which is the Davey-Stewartson equation, where $p=E\bar q$.
As well known these equations are too gauge equivalent each other[10].

4) The Myrzakulov-IX equation has form[1]
$$ iS_t+\frac{1}{2}[S,M_1S]+A_2S_x+A_1S_y = 0 \eqno(31a)$$
$$ M_2u=\frac{\alpha}{4i}tr(S[S_y,S_x]) \eqno(31b)$$
where $ \alpha,b,a  $=  consts and
$$ M_1= \alpha ^2\frac{\partial ^2}{\partial y^2}-2\alpha (b-a)\frac{\partial^2}
   {\partial x \partial y}+(a^2-2ab-b)\frac{\partial^2}{\partial x^2}; $$
$$ M_2=\alpha^2\frac{\partial^2}{\partial y^2} -\alpha(2a+1)\frac{\partial^2}
   {\partial x \partial y}+a(a+1)\frac{\partial^2}{\partial x^2},$$
$$ A_1=i\alpha\{(2ab+a+b)u_x-(2b+1)\alpha u_y\} $$
$$ A_2=i\{\alpha(2ab+a+b)u_y-(2a^2b+a^2+2ab+b)u_x\}, $$
Eqs.(31) admit the two integrable reductions. As b=0, eqs. (31)
after the some manipulations reduces to the M-VIII equation (26) and as
 $ a=b=-\frac{1}{2} $
to the Ishimori equation(29). In general we have the two integrable cases:
 the M-IXA equation as $\alpha^{2} = 1,$ the M-IXB equation
as $\alpha^{2} = -1$. We note that the M-IX equation is integrable and
admits the following Lax representation [1]
$$ \alpha \Phi_y =\frac{1}{2}[S+(2a+1)I]\Phi_x \eqno(32a) $$
$$ \Phi_t=\frac{i}{2}[S+(2b+1)I]\Phi_{xx}+\frac{i}{2}W\Phi_x \eqno(32b) $$
where $$ W_1=W-W_2=(2b+1)E+(2b-a+\frac{1}{2})SS_x+(2b+1)FS $$
$$ W_2=W-W_1=FI+\frac{1}{2}S_x+ES+\alpha SS_y $$
$$ E = -\frac{i}{2\alpha} u_x,\,\,\,  F = \frac{i}{2}(\frac{(2a+1)u_{x}}{\alpha} -
2u_{y}) $$

Hence as $b = 0$ we get the Lax refresentations of the M-VIII(26)
as $b = 0$ and for the Ishimori equation (29)$a=b=-\frac{1}{2}$. The
M-IX equation(31)  admit the different type exact solutions
(solitons, lumps, vortex-like, dromion-like and so on)[7].
As shown in [1] eqs. (31) have the following L-equivalent counterpart
$$ iq_t+M_1q+vq=0 \eqno(33a)$$
$$ ip_t-M_1p-vp=0 \eqno(33b)$$
$$ M_2v=M_1(pq) \eqno(33c)$$
where $p=E\bar q$. As well known these equations are too integrable [4]
and as in the previous case, equations (33) have the two integrable
reductions: equations(28) as $b=0$ and the Davey-Stewartson equation(30)
as $ a=b=-\frac{1}{2}. $

5) The Myrzakulov-XXII equation has form[1]
$$ -iS_t=\frac{1}{2}([S,S_y]+2iuS)_x+\frac{i}{2}V_1S_x-2ia^2 S_y \eqno(34a) $$
$$ u_x=-\vec S(\vec S_x\wedge \vec S_y) \eqno(34b)$$
$$ V_{1x}=\frac{1}{4a^2}(\vec S^2_x)_y \eqno(34c)$$
The L-equivalent of these equations are given by[1]
$$ q_t=iq_{yx}-\frac{1}{2}[(V_1q)_x-qV_2-qrq_y] $$
$$ r_t=-ir_{yx}-\frac{1}{2}[(V_1r)_x-qrr_y+rV_2] $$
$$ V_{1x}=(qr)_y $$
$$ V_{2x}=r_{yx}q-rq_{yx} $$
where $r=E\bar q$. Both these set of equations are integrable and the corressponding
Lax representations were presented in[1].
\\
\\
\\
{\bf IV. CONCLUSION}
\\
\\

In this paper using the C-approach[1] we have presented the L-equivalent
soliton equations of the Ishimori and some Myrzakulov equations. Finally
we note that using the C-approach we can see to the older problems from
the new point of view. For example, the isotropic Landau-Lifshitz equation
$$ \vec S_{t}=\vec S\wedge \vec S_{xx} $$
and the NLSE
$$ iq_t+q_{xx}+2Eq^{2}\bar q = 0 $$
are L-equivalence each other[2]. In our case $E = \pm 1$. At the same time
the 1+1 dimensional M-IV and M-XXI equations have the following L-eqivalents:\\
the 1+1 dimensional mKdV
$$ q_t+q_{xxx}+6Eq^{2}q_{x} = 0 $$
the 1+1 dimensional KdV
$$ q_t+q_{xxx}+6Eqq_{x} = 0 $$
respectively.

\end{document}